\documentclass[journal]{vgtc}                     

\onlineid{0}

\vgtccategory{Research}

\vgtcpapertype{please specify}

\title{Theory is Shapes}

\author{%
 Matthew Varona, Maryam Hedayati, Matthew Kay, and Carolina Nobre
}

\authorfooter{
\item
  Matthew Varona and Carolina Nobre are with the University of Toronto. E-mails: [varona, cnobre]@cs.toronto.edu.
\item
  Maryam Hedayati is with Princeton University. \\E-mail: mhedayati@princeton.edu.
\item
  Matthew Kay is with Northwestern University.  \\E-mail: mjskay@northwestern.edu.
}

\abstract{%
"Theory figures" are a staple of theoretical visualization research. Common shapes such as Cartesian planes and flowcharts can be used not only to explain conceptual contributions, but to think through and refine the contribution itself. Yet, theory figures tend to be limited to a set of standard shapes, limiting the creative and expressive potential of visualization theory. In this work, we explore how the shapes used in theory figures afford different understandings and explanations of their underlying phenomena. We speculate on the value of visualizing theories using more expressive configurations, such as icebergs, horseshoes, Möbius strips, and BLT sandwiches. By reflecting on figure-making’s generative role in the practice of theorizing, we conclude that theory is, in fact, shapes.}


\teaser{
  \centering
  \includegraphics[width=0.9\linewidth, alt={The iceberg of visualization theory figures.}]{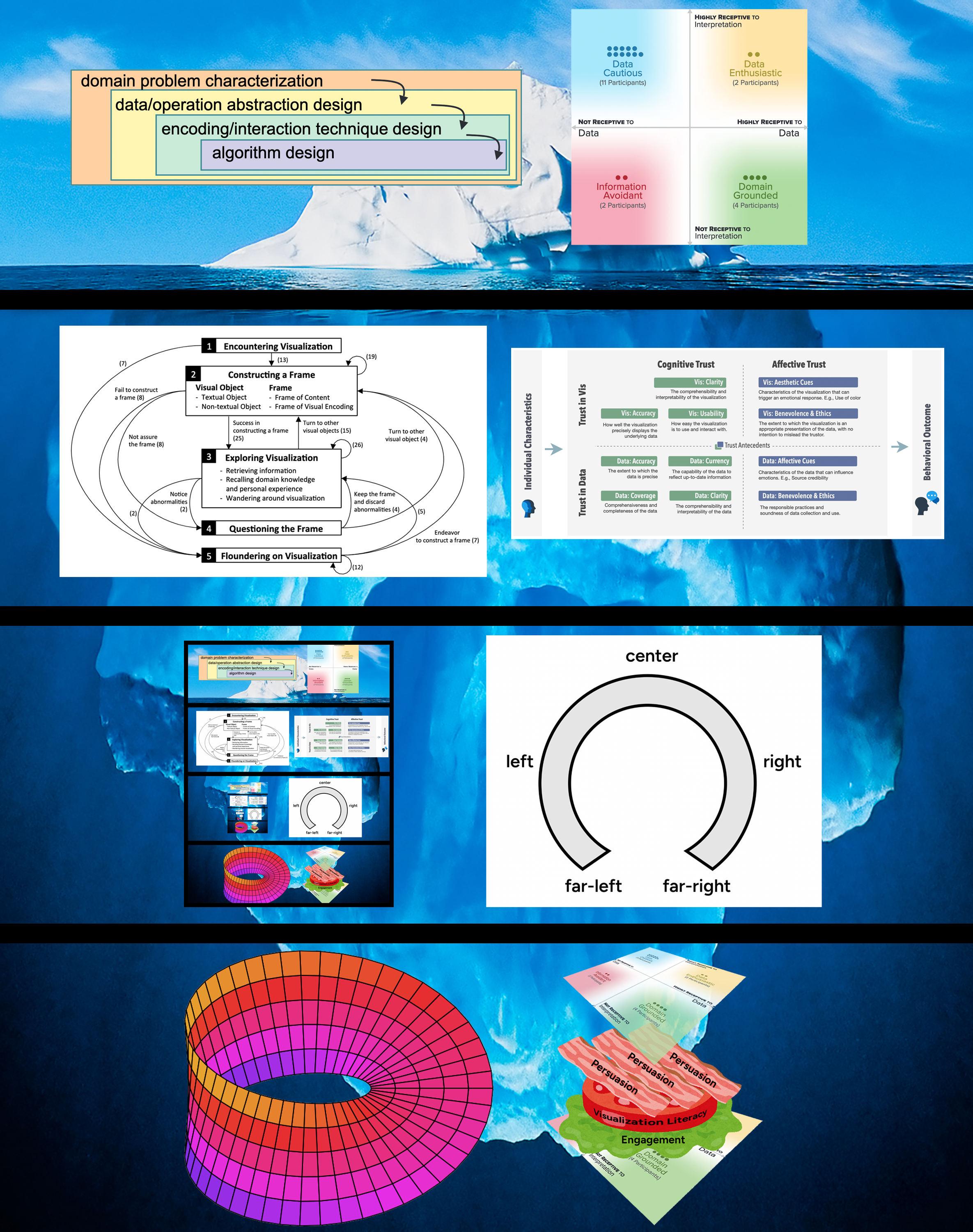}
  \caption{The iceberg of visualization theory figures.
  }
  \label{fig:teaser}
}

\renewcommand{\manuscriptnotetxt}{}

\graphicspath{{figs/}{figures/}{pictures/}{images/}{./}} 

\usepackage{xurl}
\usepackage{xcolor}

\newif\ifnotes
\notestrue 
\notesfalse 

\definecolor{todocolor}{HTML}{D90429}

\notestrue{}

\begin{document}


\maketitle
\section{Introduction}

Theory gives shape to scholarly work. It can condense messy phenomena into structured explanations, abstracting away details in order to highlight specific patterns or relationships. In visualization research, conceptual contributions like models, frameworks, and taxonomies underpin how we design systems and study users. Theory is not just textual; it is often visual, too. Figures that depict a theory's components and relationships---what we call ``theory figures''---aid researchers in both communicating and constructing theoretical contributions. Such figures are common in systematic literature reviews, qualitative research, or any other work that presents conceptual contributions. 

Well-chosen figures can do the work of an entire page of exposition \cite{PictureWorthThousand2025}. A Cartesian plane invites the reader to imagine a two-dimensional space of variation, with distinct quadrants and axes. Similarly, a flowchart can depict complex processes or dependencies at a glance. In doing the heavy lifting to explain a theory, shapes can also become synonymous with the theory itself, helping it stick in the minds of readers. When designing a figure, we are not merely illustrating a theory; we are, in a very real sense, thinking through shapes.

Yet, figures in visualization research rely heavily on a narrow canon of shapes—grids, flowcharts, matrices, and the occasional Euler diagram.\footnote{In this paper, we liberally use the term ``shapes'' to refer to the different visual forms theory can take. Strictly speaking, some of these aren't exactly shapes, but you could write an entire alt.vis paper in the time it takes to say ``different visual forms'' 59 times, which is the amount of times we say ``shape'' in this paper.} Reliance on these familiar shapes can limit the creative potential of visualization theory. What might we learn if we embraced the value of more expressive shapes, such as Möbius strips, icebergs, and BLT sandwiches? Could these shapes afford new ways of constructing, organizing, and sharing visualization theory?

In this work, we explore the landscape of theory figures in visualization research. We begin by discussing the diagrammatic affordances of four conventional shapes in theory figures—Cartesians, flowcharts, matrices, and set diagrams. We then speculate on the value of using more unique shapes, considering examples like ``The BLT Theory of Visualization Consumption''. We conclude by reflecting on how theory-figure-making can be not just a matter of communication, but a core practice of theorization.

\section{Background}

Scholars have long reflected on the visual and rhetorical forms of research itself. Prior work includes automatic generation of IEEE VIS paper titles based on common naming conventions \cite{LangetalThisNameCan2022}, as well as using computer vision to analyze the quality of a paper solely based on its layout \cite{vonBearnensquashPaperGestalt}.\footnote{The resulting system seemed to score papers higher if they had lots of large, colorful figures. Thankfully, the system doesn't look at actual content, so we suspect our paper would score fairly well based on that metric alone.} While these approaches are mainly quantitative and focus on different paper elements than ours, they share with our work a curiosity about how form shapes meaning in research. 

Our interest in theory figures also draws on the area of visual semiotics, most notably Jacques Bertin’s Semiology of Graphics \cite{BertinSemiologyGraphicsDiagrams2011}. Bertin demonstrated that graphical forms---points, lines, areas, and their arrangements---can encode information and meaning. We extend this perspective to contemporary theory figures, themselves structured graphics whose shapes influence interpretation. 

\section{Conventional Figure Shapes and Their Affordances}
The shape of a theory figure is not neutral. A figure’s geometry guides how a model is interpreted---for example, a flowchart encourages the reader to follow along possible paths, while a table invites comparison across different categories. We describe these qualities as a figure’s \textit{affordances}, adapting the term from human-computer interaction (HCI) research \cite{NormanDesignEverydayThings2013}. While affordances in HCI are about the actions possible with an interface, we use the term more loosely, extending it into the domain of conceptual and interpretive possibilities. In the context of theory figures, affordances capture how a shape invites particular ways of thinking---both for the theorist constructing the model and for the reader interpreting it.

We begin by looking at four of the most common shapes seen in theory figures: Cartesian planes, matrices, networks, and set diagrams. Each has distinct affordances that influence the structure and emphasis of the theories they depict. These affordances serve dual roles: they invite certain interpretations from a reader, and by extension, influence outcomes from the practice of theorizing. By first understanding how these familiar shapes guide our thinking, we can later begin to imagine more adventurous shapes for future theory figures. 

\subsection{Cartesian planes}
\begin{figure}[ht]
    \centering

    \begin{subfigure}[t]{0.48\columnwidth}
        \centering
        \includegraphics[height=1.2in]{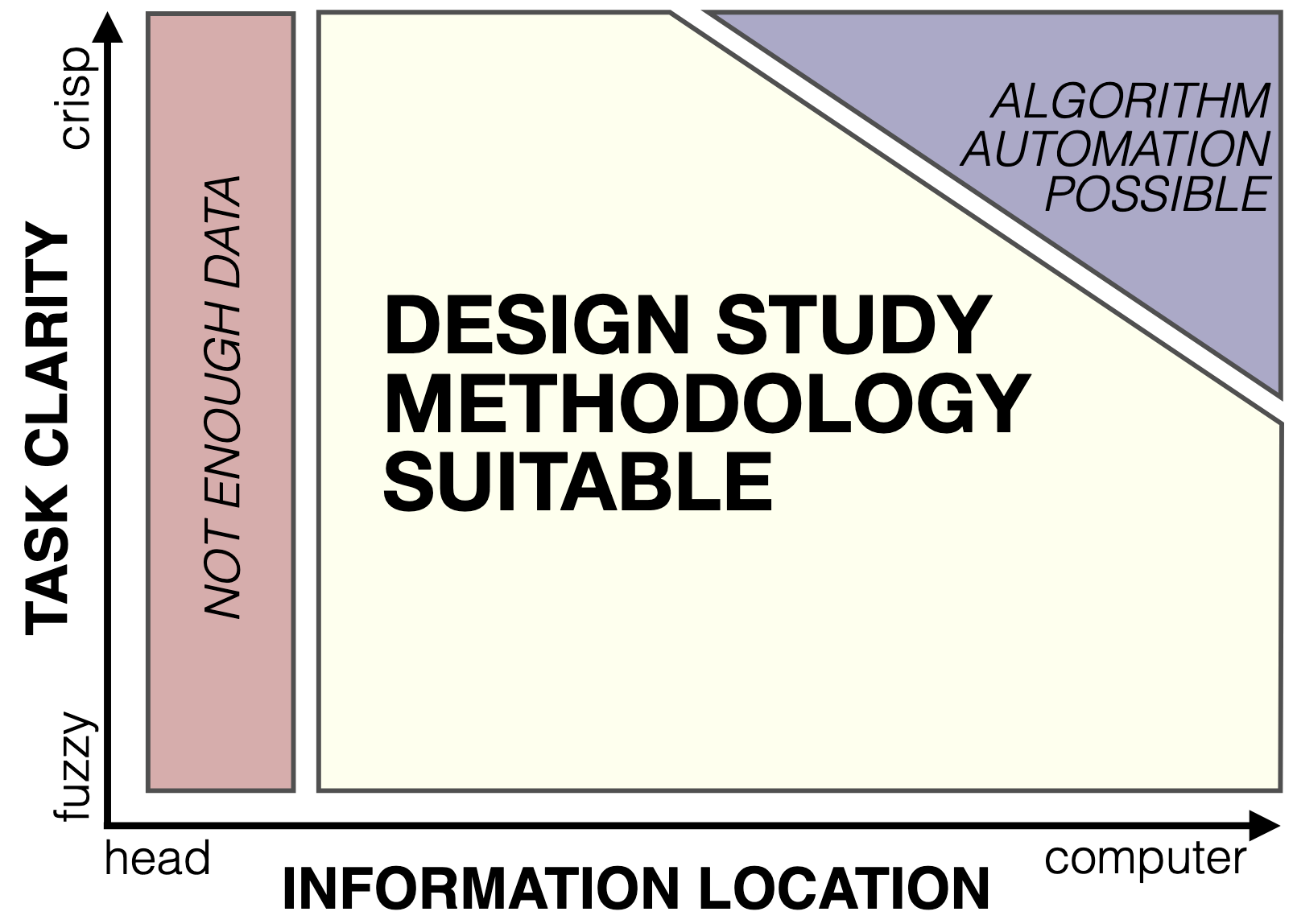}
        \caption{Space of possible design study scenarios}
        \label{fig:dsm}
    \end{subfigure}
    \hfill
    \begin{subfigure}[t]{0.48\columnwidth}
        \centering
        \includegraphics[height=1.4in]{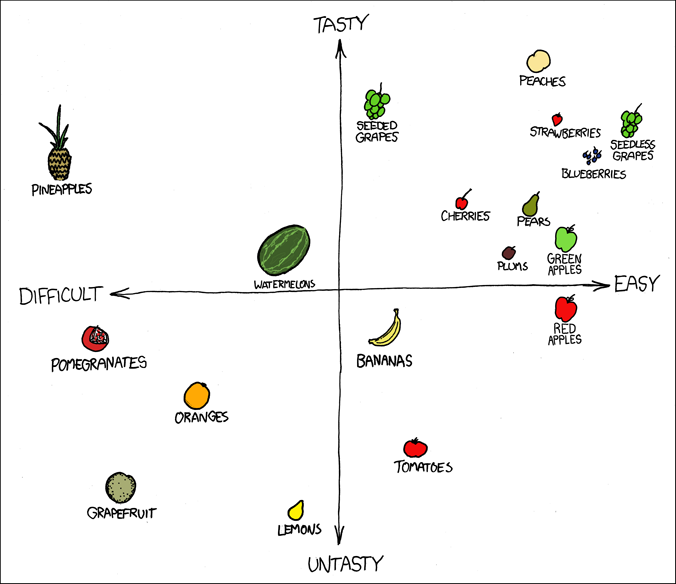}
        \caption{Space of fruits}
        \label{fig:fruit}
    \end{subfigure}
    \caption{Cartesian planes afford continuous positioning as well as discrete segmentation of their elements. In (a), Sedlmair et al. \cite{SedlmairMeyerMunznerDesignStudyMethodology2012} provide guidance on when it is suitable to conduct a design study, based on task clarity and information location. The xkcd comic in (b) organizes fruits along axes of tastiness and ease of eating \cite{xkcdFuckGrapefruit}.}
    \label{fig:combined}
\end{figure}
Cartesian planes consist of two axes that represent continuous variables. They are commonly used to map a conceptual space, situating ideas, phenomena, or actors based on their positions along two meaningful dimensions. This makes the Cartesian plane especially well-suited to theories that involve trade-offs or correlations between variables.

A core affordance of the Cartesian plane is the ability to partition. Axes can divide the plane into four quadrants, creating distinct regions based on the variables of the design space. This quadrant logic lends itself well to typologies: e.g., “high‑high,” “low‑high,” “high‑low,” and “low‑low” patterns, such as in \cref{fig:fruit} \cite{xkcdFuckGrapefruit}. This type of theory figure has seen popular use in the form of the ``political compass'', which claims to organize political views along economic (left-right) and social (authoritarian-libertarian) axes \cite{PoliticalCompass}.

In addition to quadrants, Cartesian planes afford more granular ``sub-areas'' that can represent meaningful subsets or scenarios. Sedlmair et al. \cite{SedlmairMeyerMunznerDesignStudyMethodology2012} use this technique to help researchers decide when a visualization design study is appropriate, as seen in \cref{fig:dsm}. These two types of grouping can complement or complicate each other; for example, a four-quadrant space could contain sub-areas that cross axis boundaries, showing edge cases or commonalities between quadrants. Through these different types of segmentation, Cartesian planes are able to represent both continuous and discrete elements of a theory.

While coordinate systems in figures are generally limited to 2 dimensions, we acknowledge the potential of 3-or-more-dimensional conceptual spaces as well. However, such figures may suffer from poor readability until academia breaks free from the tyranny of static paper formats \cite{DragicevicetalIncreasingTransparencyResearch2019, HeeretalLivingPapersLanguage2023}. In lieu of interactivity or animation, Cartesian planes can represent additional variables through the use of color, size, or shape, similar to scatter plots.

\subsection{Matrices and tables}
\begin{figure}[ht]
    \centering
    \begin{subfigure}[t]{0.48\columnwidth}
        \centering
        \includegraphics[height=1.5in]{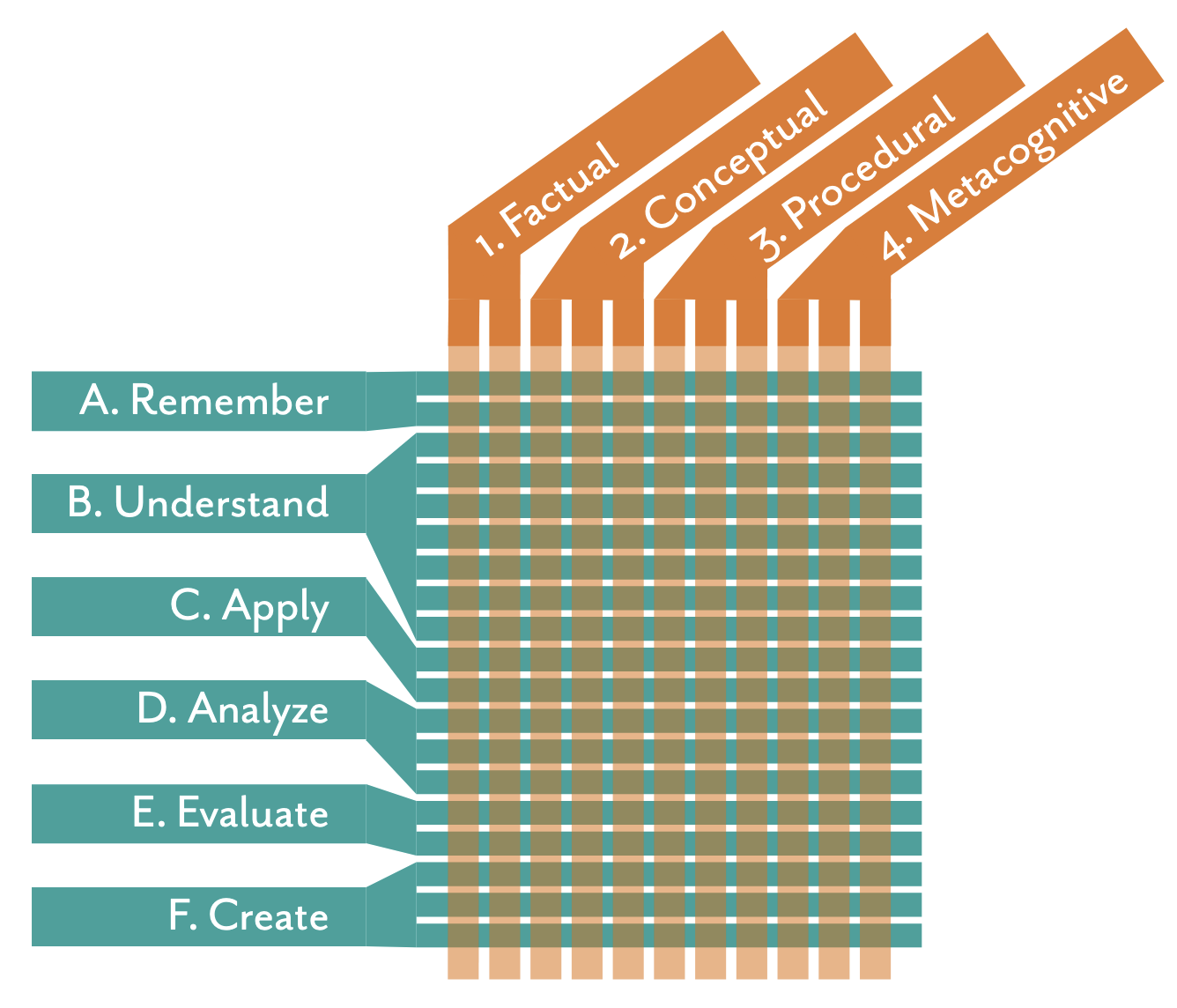}
        \caption{Matrix-based extension of Bloom's Taxonomy}
        \label{fig:bloom}
    \end{subfigure}
    \hfill
    \begin{subfigure}[t]{0.48\columnwidth}
        \centering
        \includegraphics[height=1.5in]{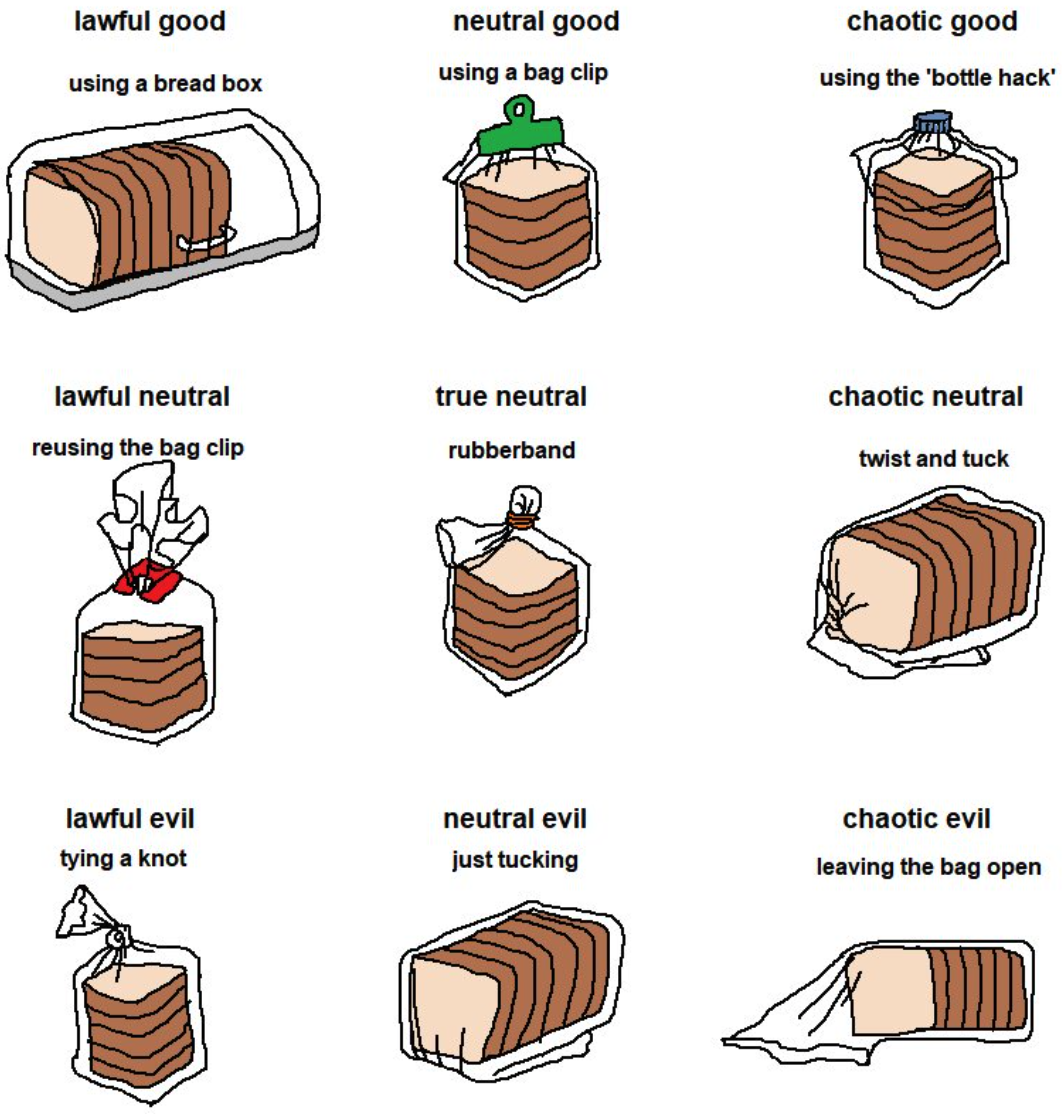}
        \caption{Bread alignment chart}
        \label{fig:breadnd}
    \end{subfigure}
    \caption{Matrices organize knowledge into a neat grid, allowing rapid comparison and scanning. In (a), Adar and Lee use a matrix to modify an existing educational taxonomy \cite{AdarLeeCommunicativeVisualizationsLearning2021}. The meme in (b) organizes different ways of storing bread based on character alignments from the game Dungeons \& Dragons \cite{AlignmentCharts2009}.}
    \label{fig:combined}
\end{figure}
Matrices, in their simplest form, are two-dimensional grids that organize information into rows and columns. Each cell represents the intersection of a specific row and column, allowing for categorical or ordinal comparisons along two dimensions. Unlike Cartesian planes, which are primarily continuous, matrices favor discrete, combinatorial structure. They invite viewers to scan, compare, and categorize, rather than interpolate.

A popular example is the alignment chart from the role-playing game Dungeons \& Dragons. Alignment charts use two axes—Lawful/Neutral/Chaotic and Good/Neutral/Evil—to explain the values and behaviors of in-game characters. The format has become ubiquitous in online culture, used to sort everything from fictional characters to bread storage methods, as seen in \cref{fig:breadnd} \cite{AlignmentCharts2009}. Keen readers may notice that matrices with ordered axes (such as alignment charts) are just subdivided Cartesian planes. While this may be true for D\&D charts, matrices generally ignore position within their cells. Subdivided Cartesian planes, however, still treat position within each region as meaningful.

One can extend the logic of matrices into hierarchical tables, where one or both axes have nested categories instead of a flat list. These are often used to show hierarchy in taxonomies, as in Adar and Lee's cognitive taxonomy for communicative visualizations \cref{fig:bloom} \cite{AdarLeeCommunicativeVisualizationsLearning2021}. Depending on how hierarchy is introduced, it is possible for the table to lose some of its matrix properties (for example, if the uniform grid is interrupted or cells are grouped).

\subsection{Networks}
\label{networks}
\begin{figure}[ht]
    \centering
    \begin{subfigure}[t]{0.49\columnwidth}
        \centering
        \includegraphics[width=\columnwidth]{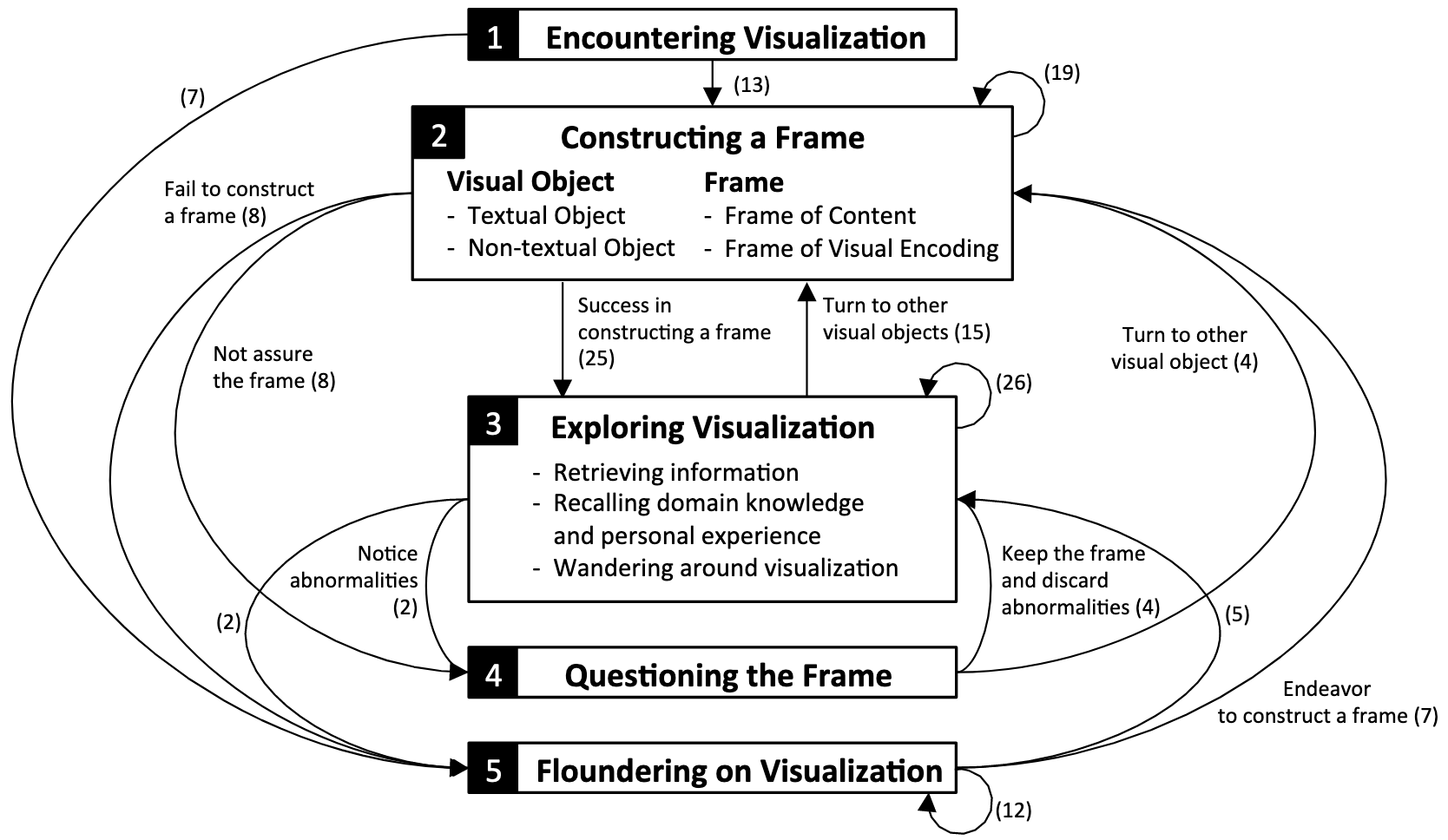}
        \caption{Novice sensemaking process}
        \label{fig:novis}
    \end{subfigure}
    \hfill
    \begin{subfigure}[t]{0.49\columnwidth}
        \centering
        \includegraphics[width=\columnwidth]{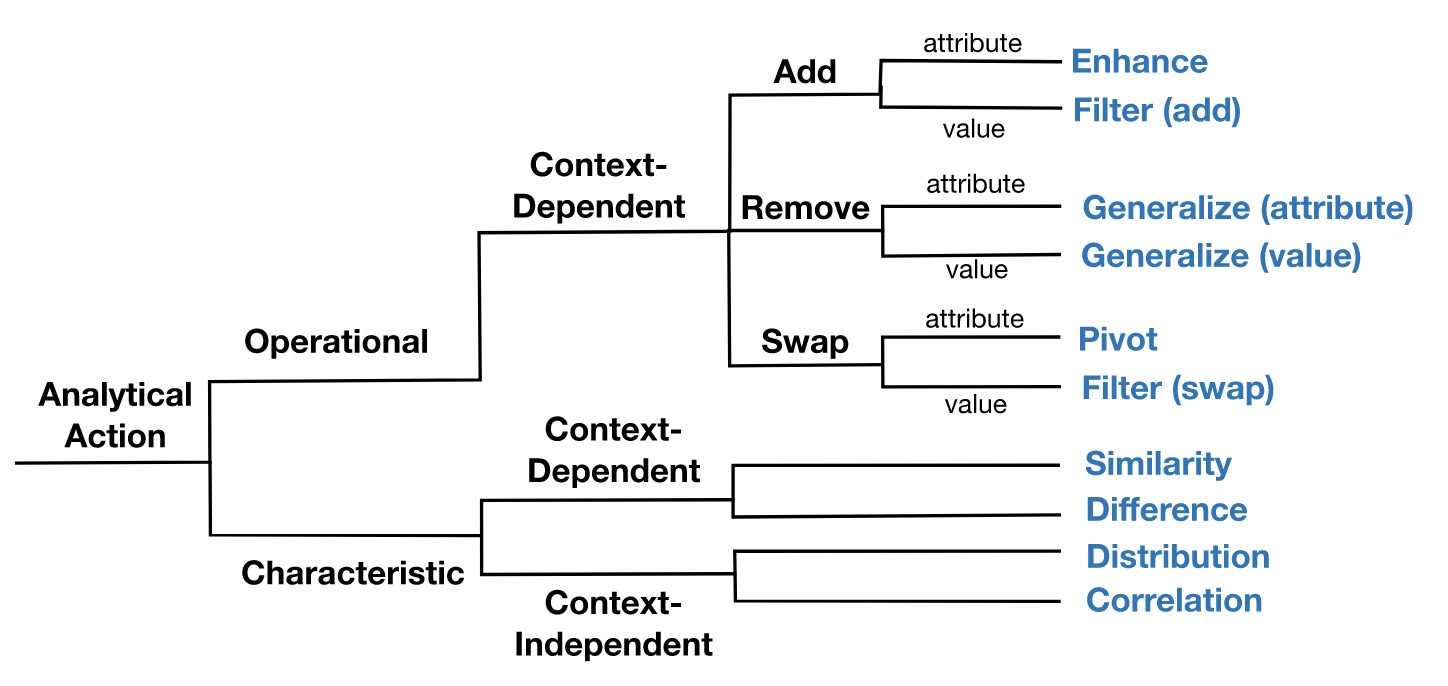}
        \caption{Hierarchy of actions in visualization recommendation}
        \label{fig:reco}
    \end{subfigure}
    \caption{Networks are suitable for representing processes, connections, or hierarchies. In (a), Lee et al. \cite{LeeetalNOVIS2016} depict the sensemaking process as a directed graph containing cycles and branches. In (b), Lee et al. show the hierarchy of actions in visualization recommendation systems.}
    \label{fig:combined}
\end{figure}

Networks consist of nodes and edges, and can surface relationships, dependencies, and flows between entities. They often occur as flowcharts or directed graphs that depict processes. Unlike Cartesian planes or matrices, which afford positioning and categorization, networks excel at showing connections and sequences. They can branch, recurse, and converge, making them an ideal choice for visualizing elements with complex relationships.

A key affordance of networks is path-following: arrows can trace progression from one node to the next. Networks also afford recursion and feedback in a way that other diagram types do not—cycles and loops can be drawn directly into the structure, making it easy to represent messy sequences. Lee et al. \cite{LeeetalNOVIS2016} use these techniques to depict how novices make sense of visualizations, which is an iterative and non-linear process (\cref{fig:novis}). 

Like other common shapes, networks can depict hierarchy in the form of trees. Lee et al. \cite{LeeetalDeconstructingCategorizationVisualization2022} use a tree structure to taxonomize the types of actions used in visualization recommendation systems (\cref{fig:reco}). An advantage of trees is that they can efficiently show many layers of hierarchy, while also letting readers trace a path from the root category to a leaf. Aside from hierarchies and processes, networks could be used to show many-to-many relationships or dependencies, although care should be taken to minimize edge crossings for readability.\footnote{Or, if you want worse readability for some reason, you can deliberately maximize edge crossings \cite{DiBartolomeoLangDunneWorstGraphLayout2022}.} 

\subsection{Set diagrams}
\begin{figure}[ht]
    \centering
    \begin{subfigure}[ht]{0.49\columnwidth}
        \centering
        \includegraphics[height=0.45in]{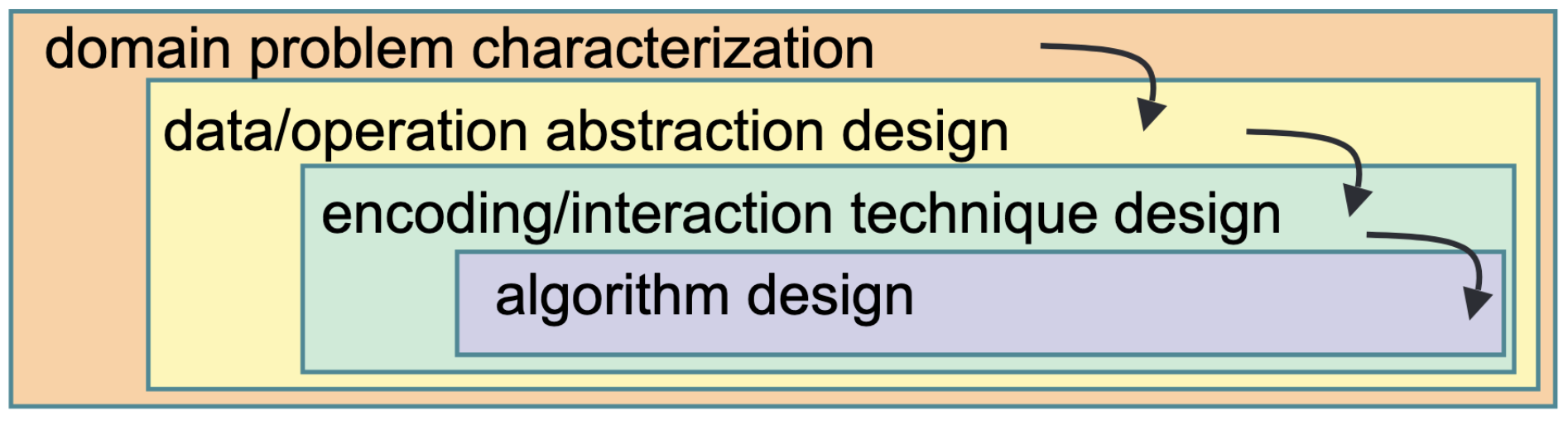}
        \caption{Nested model of visualization design}
        \label{fig:nest}
    \end{subfigure}
    \hfill
    \begin{subfigure}[ht]{0.49\columnwidth}
        \centering
        \includegraphics[height=1in]{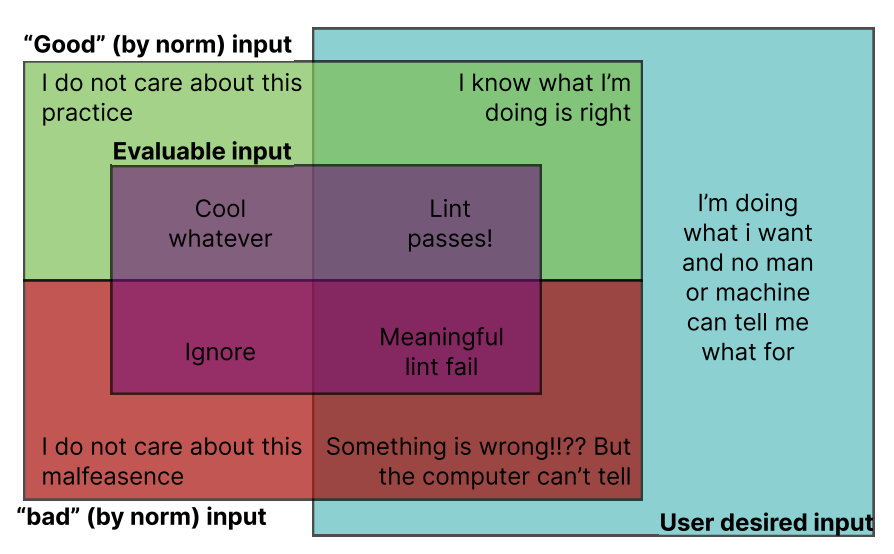}
        \caption{Input space for linting}
        \label{fig:lint}
    \end{subfigure}
    \caption{Set diagrams can show intersection, membership, and dependency. Munzner's model in (a) depicts visualization design layers as a series of nested subsets \cite{MunznerNestedModelVisualization2009}. Meanwhile in (b), Crisan and McNutt use an Euler diagram to show the relationships between good, bad, and user-desired inputs for linting \cite{CrisanMcNuttLintingPeopleExploring2025}.}
    \label{fig:combined}
\end{figure}

Set diagrams show how concepts relate to each other through overlap and nesting. The primary types of set diagrams are Euler diagrams and Venn diagrams---the main difference between the two is that Venn diagrams depict all possible relationships, while Euler diagrams can depict a relevant subset of relationships. These figures are best suited for theories about membership and shared properties. Like networks, they can also be used to convey dependency.

A well‑known example in visualization research is Munzner’s nested model of visualization design (\cref{fig:nest}) \cite{MunznerNestedModelVisualization2009}, which uses a layered, set‑like figure to show how lower‑level design decisions (such as algorithms) are contained within higher‑level concerns (like tasks and domain problems). While the nested model shows subset relationships, set diagrams are also capable of showing intersection or disjointedness. In \cref{fig:lint}, Crisan and McNutt characterize the space of inputs for linters using an Euler diagram \cite{CrisanMcNuttLintingPeopleExploring2025}.

Set diagrams afford an intuitive way to communicate inclusion and overlap: they make it easy to see which concepts belong together, where categories intersect, and which areas remain distinct. They can also hint at hierarchical relationships through nesting, while still accommodating cross‑cutting categories. However, their focus on membership means they are less effective for expressing dimensionality or continuous variation.

\subsection{Hybrid figures}
\begin{figure}
    \centering
    \includegraphics[width=.9\linewidth]{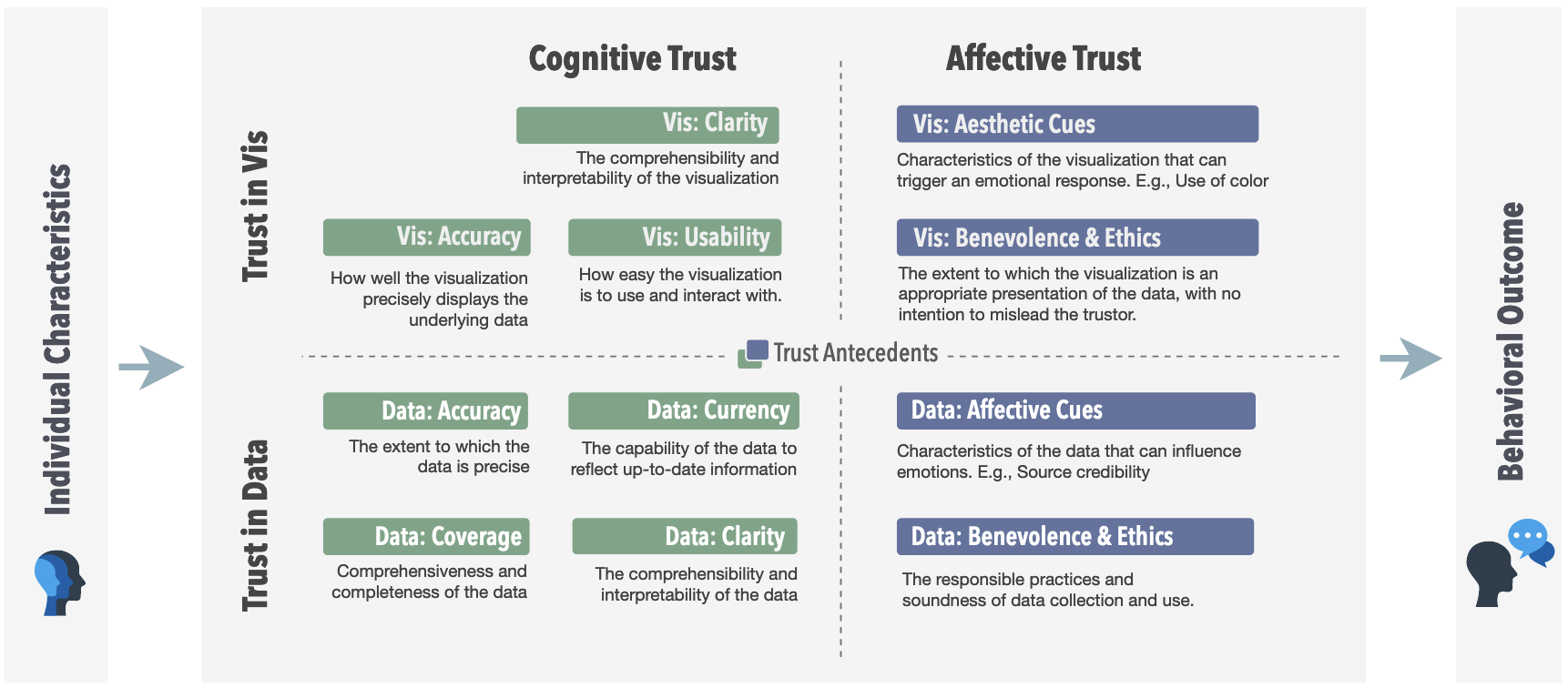}
    \caption{Elhamdadi et al.'s\cite{ElhamdadietalVistrustMultidimensionalFramework2024} framework of trust in visualizations contains both matrix and flowchart elements.}
    \label{fig:trust}
\end{figure}
Some theory figures combine multiple shapes into hybrids, capturing complexity that a single form might lack. For instance, consider Elhamdadi et al.'s framework of trust in visualizations (\cref{fig:trust}) \cite{ElhamdadietalVistrustMultidimensionalFramework2024}. This figure nests a matrix inside a flowchart, showing that trust is influenced by individual traits and affects behavioral outcomes. Hybrid figures can enable greater expressiveness, but they come with trade‑offs: complexity might make them harder to read, and an overloaded design risks diluting the clarity that simpler shapes provide. While we assert that many theories can be communicated with a single shape, hybrid figures remain a solid option when a theory spans multiple structural logics.

\section{Towards More Expressive Theory Figures}
Theory figures hold great communicative power, as demonstrated by the variety of affordances offered by conventional shapes. It follows from this logic, however, that the narrow set of currently-used theory shapes constrains the expressiveness of theory work. Notice that common shapes tend to be based on mathematical concepts; this reflects the computer science background of many visualization researchers. Mathematical precision, however, is not the only virtue of theory. How might we use more \textit{metaphorical} figures to connect with readers on a deeper level---for example, by capitalizing on our primal fear of maritime hazards, or by tantalizing hungry readers who haven't had lunch yet?

In this section, we discuss four examples of expressive shapes with the potential to describe, organize, and communicate visualization theory in novel ways. In some cases, we demonstrate the utility of these shapes by using them to propose example theories. These examples are necessarily half-formed: the intellectual labor of applying these shapes to valid, useful theories is beyond the scope of this provocation. Instead, our aim is to reveal how quickly a shape can suggest a narrative, regardless of whether or not that narrative withstands scrutiny.

\subsection{Horseshoe}
\begin{figure}[ht]
    \centering
    \begin{subfigure}[t]{.45\columnwidth}
        \centering
        \includegraphics[height=1.5in]{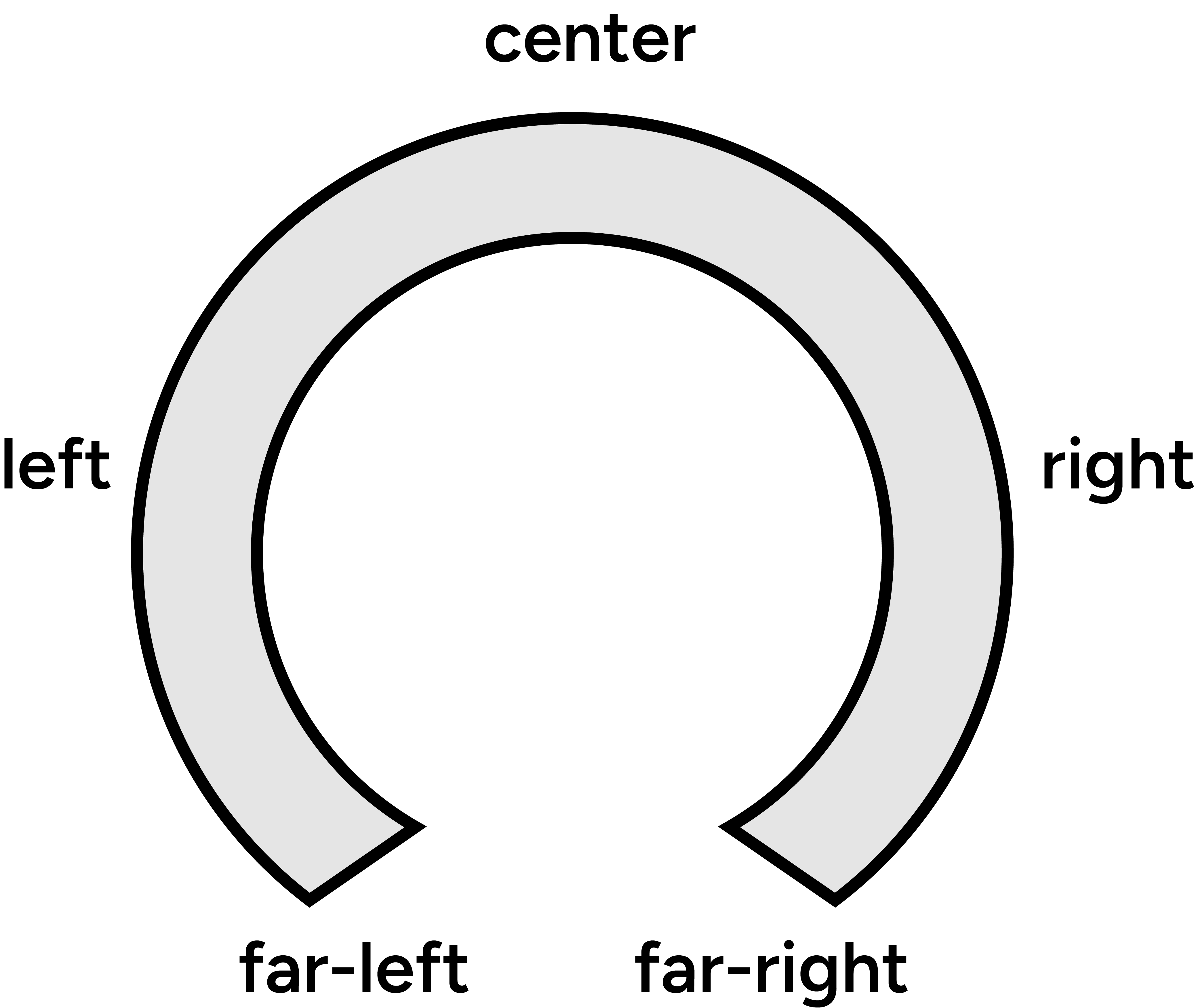}
        \caption{Political horseshoe theory}
        \label{fig:horseshoe}
    \end{subfigure}
    \hfill
    \begin{subfigure}[t]{.45\columnwidth}
        \centering
        \includegraphics[height=1.5in]{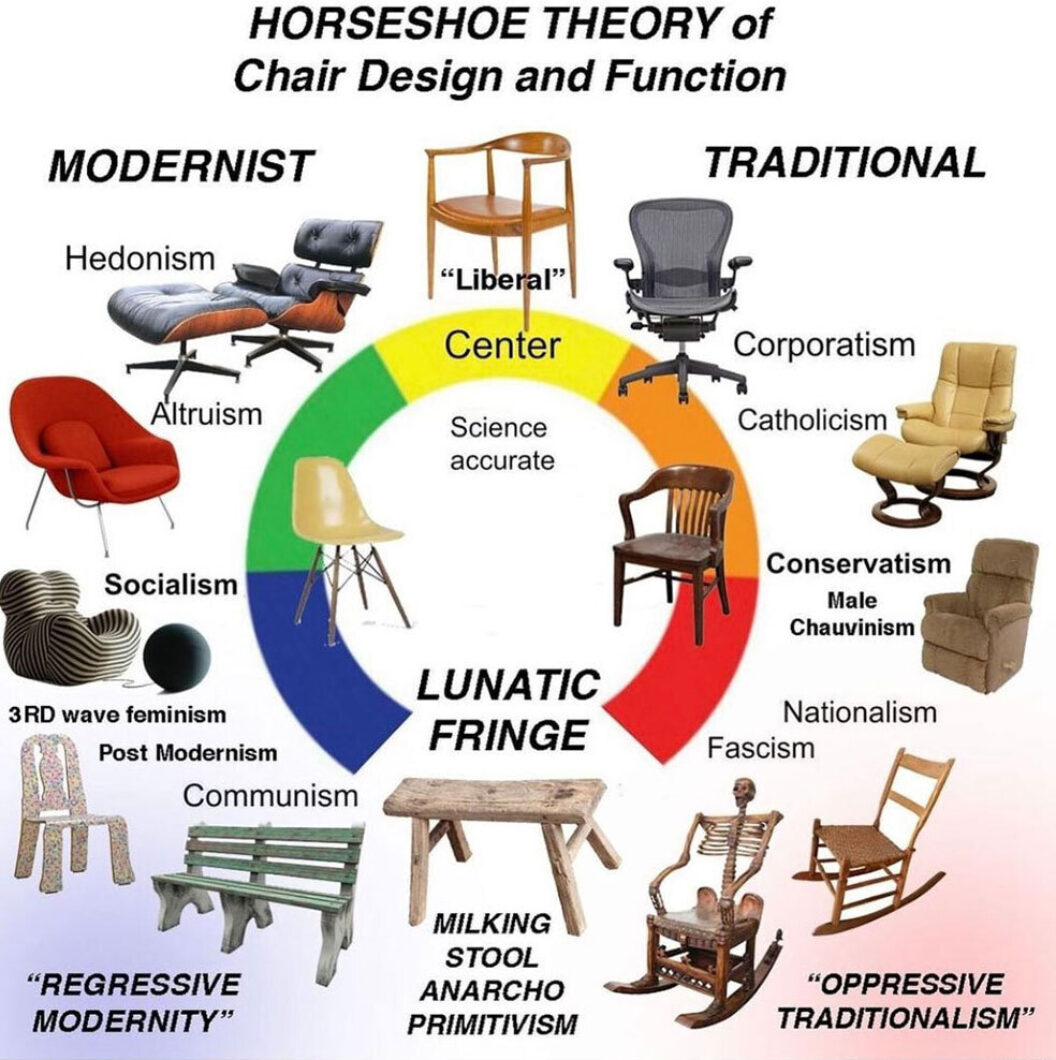}
        \caption{Horseshoe theory of chairs}
        \label{fig:chair}
    \end{subfigure}
    \caption{Horseshoes evoke parallelism between seemingly opposite points. They have seen use in politics as well as furniture design. \cite{RosenMorkeBOMBMagazineHenrike2023}}
    \label{fig:combined}
\end{figure}
A horseshoe curves an existing 1-dimensional continuum unto itself, bringing the ends of the spectrum closer together. The most popular use of this shape is in political discourse: the idea that the extreme left and right are more similar to each other than to their moderate counterparts (\cref{fig:horseshoe}). The political horseshoe theory is widely criticized by political scientists, who have pointed out that the far left and right differ vastly in their beliefs and motivations \cite{ChoatHorseshoeTheoryNonsense2017, HanelZarzecznaHaddockSharingSamePolitical2019}. Nevertheless, horseshoe theory continues to be referenced in popular discourse---proof that a compelling shape can outlive a weak theory.

Even though the political horseshoe theory is horsecrap, the shape itself has great communicative potential. It quickly conveys parallelism between two sides of a spectrum or two related processes. The ends of a horseshoe reach towards each other but do not meet\footnote{That would make it a circle.}, showing convergence while keeping end points distinct. Furthermore, the U-shape introduces a vertical axis along which another variable can be encoded. 

One could imagine folding other linear concepts---like the processes discussed in \cref{networks}---into horseshoe shapes to highlight unexpected points of similarity. Whether that is useful or misleading depends, as always, on what the shape is being asked to represent. 

\subsection{Icebergs}
The iceberg is a shape with strong cultural currency, particularly in online subcultures, where it is used to rank layers of knowledge \cite{IcebergCharts2016}. In these diagrams, the visible tip of the iceberg represents common understanding, while the submerged layers represent increasingly niche or insider knowledge. Its main affordance is communicating obscurity. The iceberg implies not just hierarchy or ranking, but concealment. It suggests that what is visible is incomplete, and that meaningful or important information lies beyond the view of the general public, as we have done in \cref{fig:teaser} through our Iceberg of Theory Figures.

Icebergs can also afford a sense of snobbery. To descend into the depths of the iceberg is to claim superior understanding, making it a solid choice for researchers hoping to signal that they are cooler than other people.\footnote{Of course, the iceberg in our teaser figure is not used this way at all. We mainly wanted to be able to make the recursion joke in the third layer.} 

Still, the iceberg is a powerful shape for theories that depend on surface/depth metaphors: symptoms vs. root causes, visible outcomes vs. invisible mechanisms. Its visual form encodes a clear conceptual split between what is seen and what is submerged, inviting the reader to descend into the underexplored depths.

A close relative of the iceberg is the tier list. Originating from video game communities, tier lists started out as a way to rank the competitive viability of game characters. They are now used to rank basically anything. While tier lists lack the visible-invisible dichotomy of icebergs, they come with their own unique affordances. Among players of fighting games, competing and winning using low-tier characters is a source of pride, showing one's drive to succeed against the odds \cite{WagarHowPickFighting2023}. Thus, tier lists could describe situations where there is tension between easy options and ``noble'' options. They are also more flexible than icebergs; tier lists can use non-competitive criteria, like which characters are most likely to believe in Santa Claus \cite{TierLists2016}.

\subsection{Möbius Strip}
\begin{figure}[ht]
    \centering
    \begin{subfigure}[t]{.45\columnwidth}
        \centering
        \includegraphics[height=1.6in]{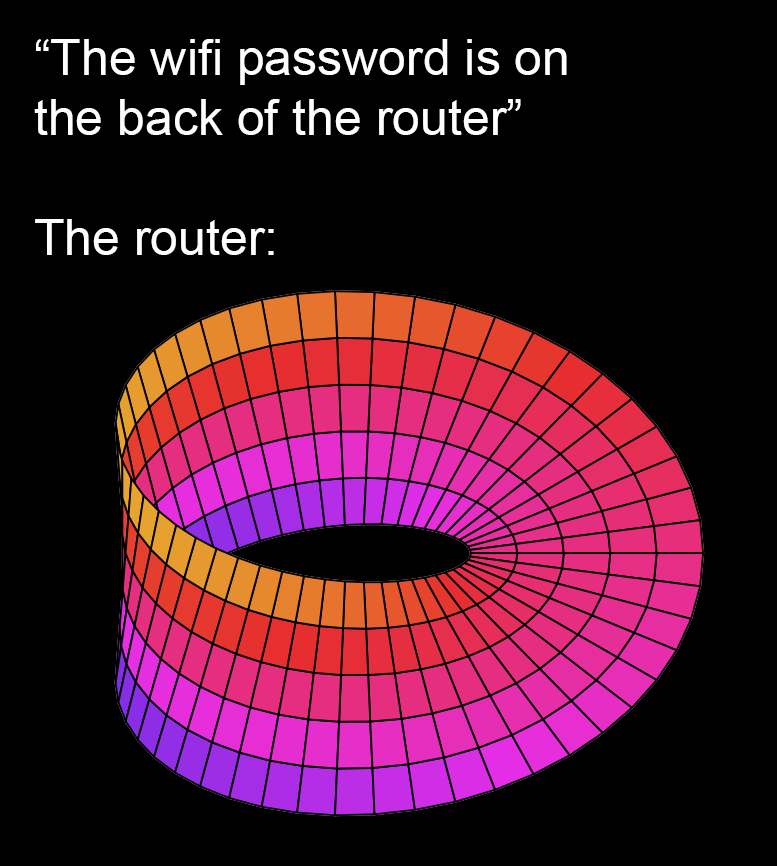}
        \caption{Möbius strips are one-sided}
        \label{fig:mobius}
    \end{subfigure}
    \hfill
    \begin{subfigure}[t]{.45\columnwidth}
        \centering
        \includegraphics[height=1.5in]{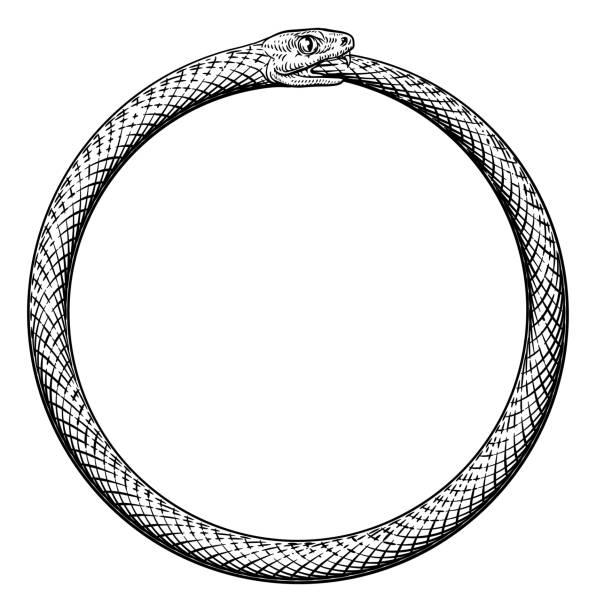}
        \caption{Ouroboros eating its own tail}
        \label{fig:ouro}
    \end{subfigure}
    \caption{A Möbius strip affords infinity and looping; the mythical ouroboros can express similar ideas, but with an air of self-defeatism.}
    \label{fig:combined}
\end{figure}

A Möbius strip is a mathematical object made by taking a strip of paper, giving it a half-twist, then joining the ends together. The resulting shape is one-sided, despite originating from a two-sided object (\cref{fig:mobius}). This non-orientable surface has various fascinating applications. Consider a ``crab canon'': a musical piece where a melody is simultaneously played forwards and backwards. By printing two halves of the melody back-to-back on a piece of paper, then turning it into a Möbius strip, one can simultaneously traverse the original and reversed melodies.\footnote{For a video demonstration of Bach's crab canon as a Möbius strip, see \href{https://www.youtube.com/watch?v=xUHQ2ybTejU}{here}.}\footnote{Technically, this depends on how the notes are being read. A Möbius strip introduces a half-twist, meaning the score ``flips'' vertically at the seam where the start and end are joined. However, if you assume that whoever is reading the music can tell up from down, it still works. For a version where vertical flipping does not matter, see \href{https://youtu.be/sToqbqP0tFk?si=NgXJmySDu3znrdUm&t=463}{here.}} Other applications include conveyor belts (ensuring that the surface receives an even distribution of wear) and computer print cartridges \cite{AlagappanTimelessJourneyMobius}.

When used for theory, the Möbius strip affords a sense of infinity and return. Tracing the surface of the strip will eventually bring you back to your starting point, but with a subtle shift---what feels like the “other side” is actually the same surface, just further along. It is well‑suited for theories that show how a process repeats with variation, or how two seemingly distinct states are actually paired aspects of the same phenomenon. Whereas the horseshoe highlights parallelism between extremes, the Möbius strip erases the boundary between them. Apparent opposites become one and the same side of a loop. 

In a similar class of shapes to the Möbius loop is the ouroboros. It is an ancient symbol of a snake devouring its own tail. While the ouroboros also represents the infinite, it frames this concept with more futility than a Möbius strip---a snake eating its own tail is doomed to repeat its destiny. As such, the ouroboros tends to be used in a more negative sense.

\subsection{BLT}
\begin{figure}
    \centering
    \includegraphics[width=0.85\linewidth]{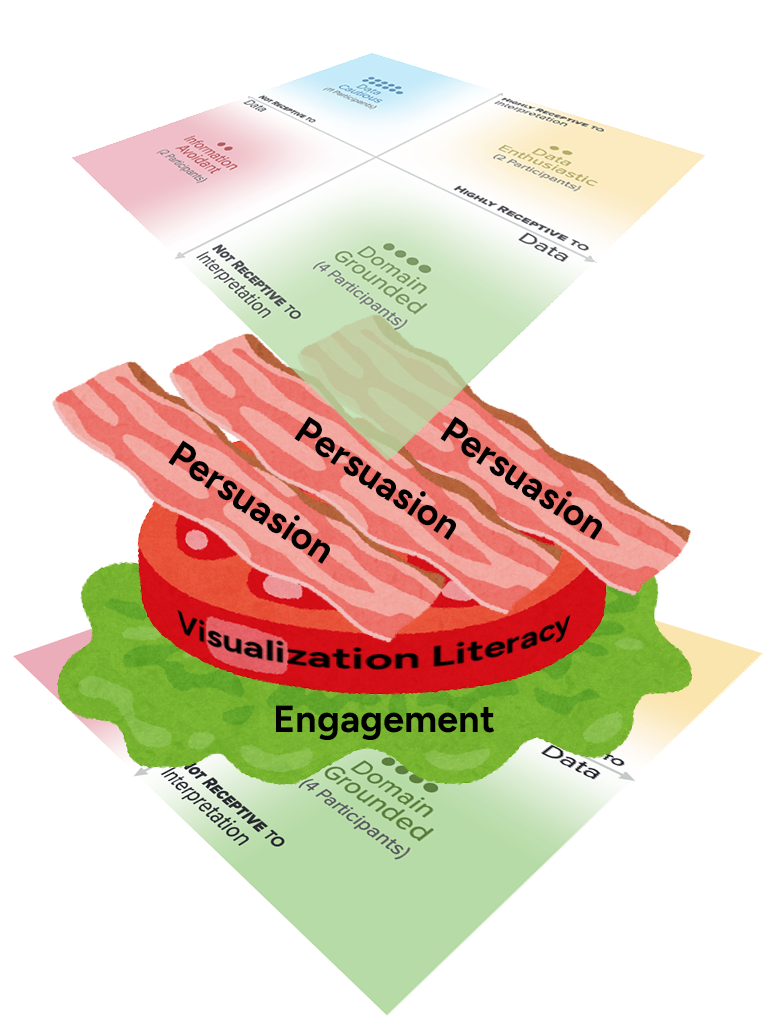}
    \caption{The BLT Theory of Visualization Consumption, depicting the interplay between visualization and reader. The sandwich consists of a visualization literacy tomato, engagement lettuce, and persuasion bacon, sandwiched between the foundational concept of information receptivity.}
    \label{fig:blt}
\end{figure}

A bacon, lettuce, and tomato (BLT) sandwich, as its name would suggest, consists of (at minimum) bacon strips \footnote{We note that our BLT sandwich uses vegan bacon.}, lettuce, and tomato slices sandwiched between two slices of bread. The multi-ingredient, layered structure of a BLT makes it well-suited to describe theoretical models where there is complex interplay between elements. In a BLT, each ingredient maintains its distinct identity while contributing to the whole. The bread provides structural foundation, while inner ingredients can represent related concepts of varying importance and function. Just as the ingredients of a BLT come together to afford deliciousness, a BLT theory is worth more than the sum of its theoretical components.

The main diagrammatic benefit of the BLT is that its ingredients can be overlaid onto any ``bread’’ (i.e. an existing theory figure). This ``bread’’ serves as a base layer that can contain any foundational theory, while the inner ingredients represent adaptations or contextual factors that modify the base theory. Consider the example in \cref{fig:blt}. Through an in-depth interview study, He et al. characterized the space of information receptivity: an individual’s openness to receiving data and interpretations of data \cite{HeetalEnthusiasticGroundedAvoidant2024}. In their work, they posit that information receptivity underpins other factors in the interplay between visualization and reader, such as literacy, engagement, and persuasion. While the original authors of the paper advocated for this in a compelling manner, we believe the point would have been driven even further home if it were expressed as the \textbf{BLT Theory of Visualization Consumption}.

As the 6th most popular sandwich in the United States \cite{GreinerWhatsAmericasFavorite}, BLTs hold immense cultural relevance. Consumers of BLTs often hold strong opinions about the ``correct” way to make the sandwich. For instance, food critic J. Kenji López-Alt asserts that ``[a] BLT is a tomato sandwich, seasoned with bacon. From this basic premise, all else follows.” \cite{Lopez-AltBestBLTSandwich} BLTs thus afford debates about theoretical primacy---which component is the true ``tomato'' of your theory, the essential element whose freshness is paramount and around which everything else is organized? Similarly, a BLT theory could be remixed with mayonnaise, avocado, or any number of auxiliary ingredients.\footnote{The authors of this paper claim no official stance on what constitutes an authentic BLT. Add ingredients at your own peril.}

We urge the community to look into the vast possibilities of food-based theories. Perhaps there could exist the complementary Grilled Cheese Theory and Tomato Soup Theory, or a time-consuming and laborious process could be captured by a Risotto Framework. The apex of theory figures might come in the form of a Möbius strip-shaped pastry---even better if it has some kind of filling.\footnote{I am getting hungry. Are you getting hungry?} 

\section{Frequently Asked Questions}

\subsection{Your new shapes suck!}
First of all, that's not a question. Second of all, no, they do not. The shapes presented here offer unique affordances that traditional figures like Cartesian planes and matrices simply cannot. A BLT captures distinct ingredients of a unified whole; a Möbius strip elegantly embodies infinity. While these shapes may not yet be common in academic publications, their expressive and analogous potential is undeniable. Our aim is to demonstrate that embracing new shapes can expand the ways we develop and communicate theory. In time, we hope to see the IEEE VIS proceedings filled with fantastical figures.

\subsection{Aren't horseshoe theory and the political compass massive oversimplifications?}
Yes! Theory is useful precisely because it abstracts detail away, and yet, poorly conceived abstraction can be misleading. Theory figures lie at the core of this tension: they wrangle some real-world concept into neat, diagrammable dimensions, in a delicate balancing act between simplicity and nuance. We maintain that political horseshoe theory's failings have little to do with the horseshoe itself compared to its uncritical development and adoption.

A tempting response to these concerns would be to add an ever-increasing number of caveats or expansions to a theory, in the hopes of defending against nitpicks from reviewers and skeptics. Sociologist Kieran Healy calls these ``nuance traps'', arguing that an obsession with nuance detracts from the abstraction necessary for theories to be useful, compelling, and even correct \cite{HealyFuckNuance2017}. As the saying goes, all models are wrong, but some are useful.\footnote{This common aphorism points out that all models are inherently wrong because they simplify complex realities. Arguably, this statement itself glosses over the range of abstraction possible within theory, although it is still a nice phrase to keep in mind. One might even say that ``all aphorisms are wrong, but some are useful''.}

\subsection{How is this related to visualization?}
Visualization theory is fundamental to how we conduct research. Our conceptual understandings of visualization processes, users, and phenomena cascade into how we conduct studies and build systems. By diving into figure design, we hope to expand the theoretical playground that visualization researchers occupy. That said, we believe the concept of diagrammatic affordances is relevant to anybody concerned with communicating and constructing theory, even beyond visualization.

\subsection{Should we \textit{always} use shapes to theorize?}
Not necessarily! Shapes are great, but there are certainly cases where they aren’t the best way to think through theory. For example, some theories might consist of mathematical relationships that are difficult to map into a coherent visual, and are better expressed in formal notation. 

That said, we maintain that thinking about shapes is more often than not a useful exercise. Theorycrafting with figures could serve a similar function as the practice of \textit{memoing} in qualitative methods. In memoing, the researcher notes down observations and ideas about low-level codes and themes, crystallizing their theoretical arguments through the process of writing \cite{BirksChapmanFrancisMemoingQualitativeResearch2008}. Even if a theory ultimately resists being drawn, the act of trying to give it a visual form can clarify relationships, reveal gaps, or surface hidden assumptions. 

\section{Conclusion}
This work proposes the idea that \textbf{theory is shapes}. We discuss the affordances offered by common shapes, demonstrating how shapes influence theory. We then describe a new set of exciting and expressive shapes with which to do theory work. However, these shapes are only the beginning. Having drawn attention to the value of shapes, we hope the community continues to sink down into the lowest layers of the theory figure iceberg.

\acknowledgments{
We wish to thank Naaz Sibia and Patrick Lee for their valuable feedback on this work. We also thank the website \href{irasutoya.com}{irasutoya.com} for providing the free clip art used to construct the BLT.
}

\bibliographystyle{abbrv-doi-hyperref-narrow}
\bibliography{references, mh_references}

\end{document}